\documentclass[twocolumn,tighten,times]{aastex631}
\usepackage{color}
\usepackage{multirow}
\usepackage{chngcntr}
\usepackage{mathtools}
\usepackage[utf8x]{inputenc}
\usepackage{perpage}
\MakePerPage{footnote}
\usepackage{lipsum}
\usepackage[graphicx]{realboxes}
\makeatletter

\newcommand{\Rmnum}[1]{\expandafter\@slowromancap\romannumeral #1@}
\makeatother

\newcommand{\sci}{Science}

\graphicspath{{./}{figures/}}

\received{\today}
\revised{tomorrow}
\accepted{the day after tomorrow}

\submitjournal{ApJ}

\shorttitle{UDGs formation}
\shortauthors{Rong et al.}

\begin{document}

\title{Gas-rich Ultra-diffuse Galaxies Are Originated from High Specific Angular Momentum}

\correspondingauthor{Yu Rong}
\email{rongyua@ustc.edu.cn}

\author{Yu Rong}
\affiliation{Department of Astronomy, University of Science and Technology of China, Hefei, Anhui 230026, China}

\author{Huijie Hu}
\affiliation{University of Chinese Academy of Sciences, Beijing 100049, China}
\affiliation{National Astronomical Observatories, Chinese Academy of Sciences, Beijing 100012, China}

\author{Min He}
\affiliation{National Astronomical Observatories, Chinese Academy of Sciences, Beijing 100012, China}

\author{Wei Du}
\affiliation{National Astronomical Observatories, Chinese Academy of Sciences, Beijing 100012, China}

\author{Qi Guo}
\affiliation{National Astronomical Observatories, Chinese Academy of Sciences, Beijing 100012, China}

\author{Hui-Yuan Wang}
\affiliation{Department of Astronomy, University of Science and Technology of China, Hefei, Anhui 230026, China}

\author{Hong-Xin Zhang}
\affiliation{Department of Astronomy, University of Science and Technology of China, Hefei, Anhui 230026, China}

\author{Houjun Mo}
\affiliation{Department of Astronomy, University of Massachusetts Amherst, MA 01003, USA}

\begin{abstract}
	
	Ultra-diffuse galaxies, characterized by comparable effective radii to the Milky Way but possessing 100-1,000 times fewer stars, offer a unique opportunity to garner novel insights into the mechanisms governing galaxy formation. Nevertheless, the existing corpus of observational and simulation studies has not yet yielded a definitive constraint or comprehensive consensus on the formation mechanisms underlying ultra-diffuse galaxies. In this study, we delve into the properties of ultra-diffuse galaxies enriched with neutral hydrogen using a semi-analytic method, with the explicit aim of constraining existing ultra-diffuse galaxy formation models. We find that the gas-rich ultra-diffuse galaxies are statistically not failed $L^{\star}$ galaxies nor dark matter deficient galaxies. In statistical terms, these ultra-diffuse galaxies exhibit comparable halo concentration, but higher baryonic mass fraction, as well as higher stellar and gas specific angular momentum, in comparison to typical dwarf galaxy counterparts. Our analysis unveils that higher gas specific angular momentum serves as the underlying factor elucidating the observed heightened baryonic mass fractions, diminished star formation efficiency, expanded stellar disk sizes, and reduced stellar densities in ultra-diffuse galaxies. Our findings make significant contributions to advancing our knowledge of ultra-diffuse galaxy formation and shed light on the intricate interplay between gas dynamics and the evolution of galaxies.

\end{abstract}



\section{Introduction} \label{sec:1}

Ultra-diffuse galaxies (UDGs) \citep{vanDokkum15} epitomize a distinctive category of galactic entities distinguished by stellar masses akin to those of traditional dwarf galaxies, yet manifesting effective radii akin to that of the Milky Way. They currently occupy a pivotal position in the domain of galactic inquiry. UDGs are characterized by their exceedingly low surface brightness and an extensive stellar disk structure, resulting in a marked deviation in their distribution on the stellar mass versus effective radius plot compared to typical dwarf and massive galaxies \citep{vanDokkum15,Leisman17,Rong20}. Within the framework of the Lambda Cold Dark Matter ($\Lambda$CDM) paradigm, prevailing galaxy evolution models generally presuppose a relatively uniform correlation between a galaxy's stellar mass and its effective radius \citep{Guo11}. Consequently, the distinctive distribution of UDGs implies a potential substantial disparity in their formation mechanism compared to typical galaxies, thus positioning UDGs as a unique vantage point for scrutinizing established galaxy evolution models and comprehending the galaxy formation process. While multiple theoretical models have been posited to elucidate the genesis of UDGs \citep{Rong17a,DiCintio17,Carleton19,Wrigts21,Grishin21}, a consensus remains elusive.

UDGs are widespread across galaxy clusters, groups, and low-density environments. Those located in high-density regions, such as galaxy clusters, are likely enveloped by substantial dark matter halos, suggesting that UDGs in dense locales may represent ``failed'' L$^{\star}$ galaxies (FLG) \citep{vanDokkum15}, with dark matter halo masses commensurate to that of the Milky Way, despite their stellar masses resembling typical dwarf galaxies. Notably, the largest UDG in the Coma cluster, DF44, has been ascertained \citep{vanDokkum16} to harbor a dark matter halo mass of $10^{12}\ \rm{M_{\odot}}$. However, subsequent statistical analyses of UDG samples, encompassing the number/mass of globular clusters and weak lensing, indicate that the majority of UDGs in galaxy clusters possess masses akin to typical dwarf galaxies ($\sim 10^{11}\ \rm{M_{\odot}}$), markedly diverging from the mass of the Milky Way \citep{Sifon18,Peng16}. Additionally, investigations into the dynamics of globular clusters in two UDGs (DF2 and DF4) within the NGC~1052 galaxy group suggest that these UDGs exhibit minimal dark matter ($\lesssim 10^7\ \rm{M_{\odot}}$) \citep{vanDokkum18,vanDokkum19}, indicating that certain UDGs may even represent dark matter-deficient dwarf galaxies (DMDD). The wide range of masses spanning from FLG to DMDD underscores the significant divergence in the formation and evolution of UDGs from typical galaxies, potentially challenging established galaxy evolution models. 

Returning to the question of the formation mechanism of UDGs, theoretically, disregarding the gravitational influence of baryonic matter, we can express the scale length and central surface density of the stellar disk formed at the core of a dark matter halo as follows \citep{Mo98},
\begin{equation} R_{\rm{\star,d}}= \frac{J_{\rm{\star}}}{2M_{\rm{\star}} V_{\rm{c}}}=\frac{S_{\star}}{2V_{\rm{c}}} \label{eq1} \end{equation}
\begin{equation} \Sigma_{\star,\rm{0}}= \frac{J_{\star}}{4\pi V_{\rm{c}}R_{\star,\rm{d}}^3}= \frac{2M_{\star}V_{\rm{c}}^2}{\pi S_{\star}^2} \label{eq2} \end{equation}
Here, $V_{\rm{c}}$ represents the circular velocity of the halo, and $M_{\rm{\star}}$, $J_{\rm{\star}}$, as well as $S_{\star}\equiv J_{\star}/M_{\star}$ characterize the mass, angular momentum, and specific angular momentum of the stellar disk, respectively. Therefore, the extended stellar distribution and low surface brightness observed in UDGs can potentially be elucidated through the following three conjectures:

(\Rmnum{1}). {\textbf{Higher $S_{\star}$}}: 
Previous cosmological hydrodynamical simulations \citep{Benavides23,Liao19} and $N$-body simulations employing semi-analytic galaxy formation models \citep{Rong17a,Amorisco16} lend support to a high-spin UDG formation scenario. These simulations posit that UDGs acquire augmented specific angular momenta from their high-spin halos. Another proposed model, involving the merging of dwarf galaxies \citep{Wrigts21}, postulates the transient amplification of descendant halos' spin during merging events to reproduce UDGs. Alternatively, UDGs may have experienced the accretion of circumgalactic medium with elevated spins while maintaining a normal halo angular momentum $J_{\rm{h}}$ comparable to typical dwarf galaxies  \citep{Posti18,Pina20}. This circumgalactic medium subsequently cools, fostering star formation and converting the heightened specific angular momentum into the stellar disk. Both scenarios, characterized by an increased halo spin and enhanced conversion efficiency $j_{\star}\equiv J_{\star}/J_{\rm{h}}$, consistently yield greater specific angular momenta $S_{\star}$ for the stellar constituents within UDGs.

Several observational inquiries into the internal kinematics of UDGs have indicated that the specific angular momentum of their stellar disks may parallel that of typical dwarf galaxies, or even suggested the absence of significant rotational motion in UDGs \citep[e.g.][]{Chilingarian19,vanDokkum19b}. Consequently, these findings have led researchers to speculate that UDGs may not have originated from dark matter halos characterized by high spin, nor do they exhibit high-spin stellar disks. It is pertinent to note, however, that these observations have predominantly focused on UDGs in high-density environments, where these UDGs have already undergone tidal heating from massive galaxies \citep[e.g.,][]{Rong20b,Pina19}, potentially diminishing their original spin, even if it was initially substantial. Nevertheless, these observational inquiries do not preclude the possibility that UDGs in low-density regions may have emerged from higher spin.

Furthermore, certain hydrodynamic simulations (e.g., NIHAO and FIRE) have suggested that the spin of UDGs is akin to that of typical dwarf galaxies \citep[e.g.,][]{DiCintio17,Chan18,Cardona-Barrero20}. However, it is important to note that these simulations are not comprehensive cosmological simulations, but rather zoom-in simulations involving a limited number of samples, and as such, lack statistical significance. In contrast, comprehensive cosmological hydrodynamic simulations such as the Illustris-TNG have indeed demonstrated that the spin statistics of UDGs exceed those of typical dwarf galaxies \citep{Benavides23}. Consequently, the hypothesis that high spin represents a plausible mechanism for the formation of UDGs, particularly in low-density regions, emerges as a compelling and potentially the most robust model of formation to date.

(\Rmnum{2}). {\textbf{Elevated $V_{\rm{c}}$ and diminished $M_{\star}$}}: The failed L$^\star$ formation model \citep{vanDokkum15} for UDGs posits the coexistence of massive host halos on par with the Milky Way \citep[resulting in an intensified $V_{\rm{c}}$ akin to that of the Milky Way, given the relationship $M_{\rm{vir}}\propto V_{\rm{c}}^3$;][]{Mo98}, alongside reduced stellar masses resembling those of typical dwarf galaxies. This attenuation in stellar mass could potentially arise from the influences of tidal interactions or ram pressure effects during the nascent stages of the Universe \citep{Yozin15}. 

(\Rmnum{3}). {\textbf{Change of gravitational potential}}: Equations (\ref{eq1}) and (\ref{eq2}) hold true only if the gravitational effects of galactic baryonic matter and the environment can be neglected \citep{Mo98}. In the cases of the outflow model \citep{Chan18,DiCintio17}, tidal interaction model \citep{Carleton19,Rong20b,Jiang19}, or environmental stripping model \citep{Grishin21} for UDG formation, the stellar orbits in UDGs expand due to significant changes in gravitational potentials.

We have systematically categorized the existing UDG formation models into these three possibilities. To ascertain the decisive factor among the aforementioned possibilities for UDG formation, in this investigation we adopt a rigorous semi-analytic approach and analyze the single-dish neutral hydrogen (HI) and optical photometric data from the Arecibo Legacy Fast Alfa Survey (ALFALFA $\alpha.100$) \citep{Giovanelli05,Haynes18} and Twelfth Data Release of the Sloan Digital Sky Survey (SDSS DR12) \citep{Alam15} to scrutinize the baryonic fractions and kinematics of HI-bearing UDGs. In section~\ref{sec:2}, we introduce the UDG samples studied in this work. In section~\ref{sec:3}, we investigate and compare the properties of gas-rich UDGs and typical dwarf counterparts, which imply the formation mechanism of UDGs. We discuss and summarize our results in section~\ref{sec:4}. In this paper, we use ``$\log$'' to represent ``$\log_{10}$''.

\section{Gas-rich UDGs in observation}\label{sec:2}

\subsection{Sample selection}\label{sec:2.1}

The UDG sample is identified from the cross-matched catalog of ALFALFA and SDSS. For ALFALFA galaxies with optical counterparts, we meticulously employ the \textsc{SExtractor} software \citep{Bertin1996} to precisely measure the $g$ and $i$-band `mag\_auto' magnitudes, effective radius $R_{\star,\rm{e}}$, as well as apparent axis ratio of each galaxy using the method proposed by Du et al. (2015) \citep{Du15}. To derive the absolute magnitudes $M_g$ and $M_i$ for galaxies with varying axis ratios and colors, we adopt the methodology outlined in Durbala et al. (2020) \citep{Durbala20} to mitigate the influence of internal dust extinctions within the galaxies. Furthermore, we utilize the $i$-band absolute magnitudes $M_i$ and $g-i$ colors to estimate galactic stellar masses $M_\star$, employing the mass-to-light ratios $\log(M_{\star}/L_i) = 0.70(g-i)-0.68-0.057$ \citep{Taylor11}, where -0.057 is a correction factor accounting for a Kroupa initial mass function \citep{Kroupa02,Herrmann16}. Subsequently, we estimate the mean stellar surface density within the effective radius as $\langle\Sigma_{\star}\rangle_{\rm{e}} \simeq M_\star/2\pi R_{\star,\rm{e}}^2$.

\begin{figure}
	\centerline{ \includegraphics[width=\columnwidth]{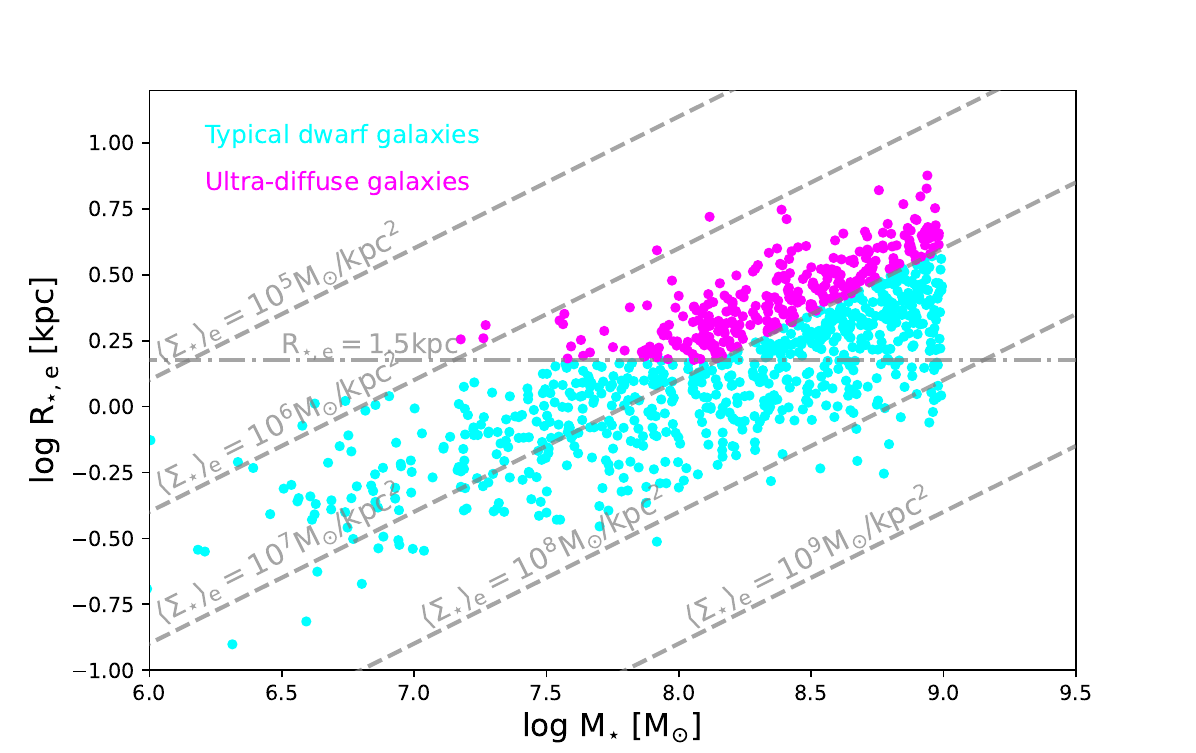} }
 	 \caption{{\bf Stellar mass $M_{\star}$ vs. effective radius $R_{\star,\rm{e}}$ plot} for UDGs (depicted by magenta circles) and typical dwarf counterparts (depicted by cyan circles).
         }
	  \label{UDG_ms}
\end{figure}

For the UDG selection, we adopt a criterion of $\langle\Sigma_{\star}\rangle_{\rm{e}}\leq 10^{7}\ {\rm M_{\odot}/kpc^{2}}$, in conjunction with $R_{\rm{e}}>1.5\ \rm kpc$ and $M_{\star}<10^9\ \rm M_{\odot}$, as shown in Fig.~\ref{UDG_ms}. This selection criterion is slightly different from the conventional criterion of identifying UDGs, which relies on surface brightness, $R_{\rm{e}}$, and $M_{\star}$. This updated approach addresses the challenge posed by varying mass-to-light ratios among different UDGs and ensures a more unbiased selection of UDG samples, particularly for galaxies exhibiting abundant HI gas, ongoing star formation, and low stellar mass-to-light ratios. In contrast, the traditional approach tends to favor UDGs with redder colors. For the UDGs in galaxy clusters and groups, with a typical color of $g-r\sim 0.6$ \citep{Rong17a} and corresponding $r$-band mass-to-light ratio of approximately ${\rm{log}}(M_{\star}/L_{\star})\sim 0.202$ \citep{Bell03}, $\langle\Sigma_{\star}\rangle_{\rm{e}}\leq 10^{7}\ {\rm M_{\odot}/kpc^{2}}$ corresponds to a surface brightness threshold of $\langle\mu_{\star}\rangle_{\rm{e}}\geq 24$~mag/arcsec$^2$ in the $r$-band, which is consistent with the selection criterion used in the work of \cite{vanDokkum15} and \cite{vanderBurg16}.

To ensure the accuracy of UDG selection and avoid potential errors arising from internal dust extinctions, galaxies with optical apparent axis ratios $b/a<0.2$ are excluded. UDGs with $b/a>0.7$ or low HI signal-to-noise ratios (SNR$<10$) are also eliminated due to significant uncertainties in estimating circular velocities. Additionally, UDGs with contaminated photometries, suspicious HI spectra, pronounced irregular optical images, or galaxy companions within a radial velocity difference $\Delta v<500$~km/s and a projected radius $R_{\rm{p}}<$7~arcmin (twice the beam size of Arecibo) are excluded, as such companions could potentially interact with UDGs or introduce inaccuracies in estimating HI masses from the HI spectra.

Consequently, a sample of 321 UDGs is obtained. The stellar masses of our UDG sample range from approximately $10^7$ to $10^9$ $\rm M_{\odot}$. To ensure a fair comparison of UDG properties with typical dwarf galaxies, galaxies with $M_\star\leq 10^9\ \rm M_{\odot}$ but $\langle\Sigma_\star\rangle_{\rm{e}}>10^7\ {\rm M_{\odot}/kpc^{2}}$ or $R_{\rm{\star,e}}\leq 1.5\ \rm kpc$ are selected as the sample of typical dwarf counterparts. 

It is noteworthy that a subset of ALFALFA galaxies lacking optical counterparts or exhibiting exceedingly faint optical signatures, commonly referred to as ``dark galaxies'' \citep{Disney76,Janowiecki15}, may fulfill the selection criteria based on effective radii and surface densities typically associated with UDGs. Nevertheless, multiple investigations have revealed that these dark galaxies are susceptible to tidal interactions \citep{Roman21,Duc08}, rendering them non-equilibrium systems. Consequently, our study excludes these entities due to the inherent challenge in accurately estimating their circular velocities and halo masses.

\subsection{Circular velocity and halo mass estimation}\label{sec:2.2}

In this study, we estimate the circular velocities $V_{\rm{c}}$ and halo masses $M_{\rm{vir}}$ for the selected UDGs and typical dwarf counterparts. 

In the absence of resolved HI data, we employ the optical $b/a$ ratio to estimate the HI disk inclination $\phi$, using $\sin\phi=\sqrt{(1-(b/a)^2)/(1-q_0^2)}$, where $q_0\sim 0.2$ \citep{Tully09,Giovanelli97,Li21} represents the intrinsic thickness of a galaxy. The circular velocity, $V_{\rm{c}}=W_{20}/2/\sin\phi$ ($W_{20}$ being the 20\% peak width of the HI line after correction for instrumental broadening \citep{Haynes18,Guo20,Hu23}), is consequently computed. We note that $W_{20}$ has been proved to be a good indicator of the asymmetric drift-corrected circular velocity using the kinematics maps of dwarf galaxies in the LITTLE THINGS \citep{Hunter12} by \cite{Guo20}. 

We introduce the HI radius $r_{\rm{HI}}$, which corresponds to the radius at which the HI surface density attains $1\ \rm M_{\odot}\rm{pc^{-2}}$. The estimation of $r_{\rm{HI}}$ is facilitated by the tight correlation observed between $r_{\rm{HI}}$ and HI mass $M_{\rm{HI}}$, as inferred from empirical observations: $\log r_{\rm{HI}}=0.51\log M_{\rm{HI}}-3.59$ \citep{Wang16,Gault21}. 

We adopt the method proposed by \cite{Guo20} to estimate the halo mass of each galaxy. Assuming a spherically symmetric dark matter halo model, we calculate the dynamical mass enclosed within the HI radius $r_{\rm HI}$,
\begin{equation}
 M_{\rm dy} (<r_{\rm HI}) =V_{\rm c}^2 r_{\rm HI}/G,
 \label{dm}
\end{equation}
where $G$ represents the gravitational constant. The halo mass $M_{\rm{vir}}$ is then estimated by assuming a Burkert dark matter profile \citep{Burkert95}, and utilizing the equations,
 \begin{equation}
\begin{aligned}
 & {M_{\rm dy}(<r_{\rm HI})}-M_{\rm{bar}} =  \int_0^{r_{\rm HI}}4\pi r^2\rho(r)\mathrm{d}r \\
 & =  2\pi \rho_{0} R_{0}^3[\ln (1+\frac{r_{\rm{HI}}}{R_{0}})+0.5\ln(1+\frac{r_{\rm{HI}}^2}{R_{0}^2}) -{\rm arctan}(\frac{r_{\rm{HI}}}{R_{0}})],
\end{aligned}
  \label{nfw}
  \end{equation}
and
\begin{equation}
	\begin{aligned}
& M_{\rm{vir}} = \int_0^{R_{\rm vir}}4\pi r^2\rho(r)\mathrm{d}r \\
& =  2\pi \rho_{0} R_{0}^3[\ln (1+\frac{R_{\rm vir}}{R_{0}})+0.5\ln(1+\frac{R_{\rm vir}^2}{R_{0}^2}) -{\rm arctan}(\frac{R_{\rm{vir}}}{R_{0}})],
\end{aligned}
\label{nfw_m}
\end{equation}
where the total baryonic mass for each galaxy is estimated as $M_{\rm{bar}}\simeq M_\star+1.33M_{\rm{HI}}$ \citep{Planck20}, the virial radius $R_{\rm vir}$ encloses a mean density of 200 times the critical value, $R_{0}$ and $\rho_{0}$ are free parameters denoting the core of the halo, and $R_0$ is related to $M_{\rm{vir}}$ as \citep{Salucci07},
\begin{equation}
\log [(R_0/{\rm kpc})] = 0.66-0.58(\log[M_{\rm{vir}}/10^{11} \rm M_{\odot}]).
\label{nfw_c}
\end{equation}
The halo concentration is evaluated as $c = R_{\rm{vir}}/R_0$.


\section{Properties of HI-bearing UDGs}\label{sec:3}

\subsection{Baryonic Tully-Fisher relation of UDGs}\label{sec:3.1}

\begin{figure}
	\centerline{ \includegraphics[width=\columnwidth]{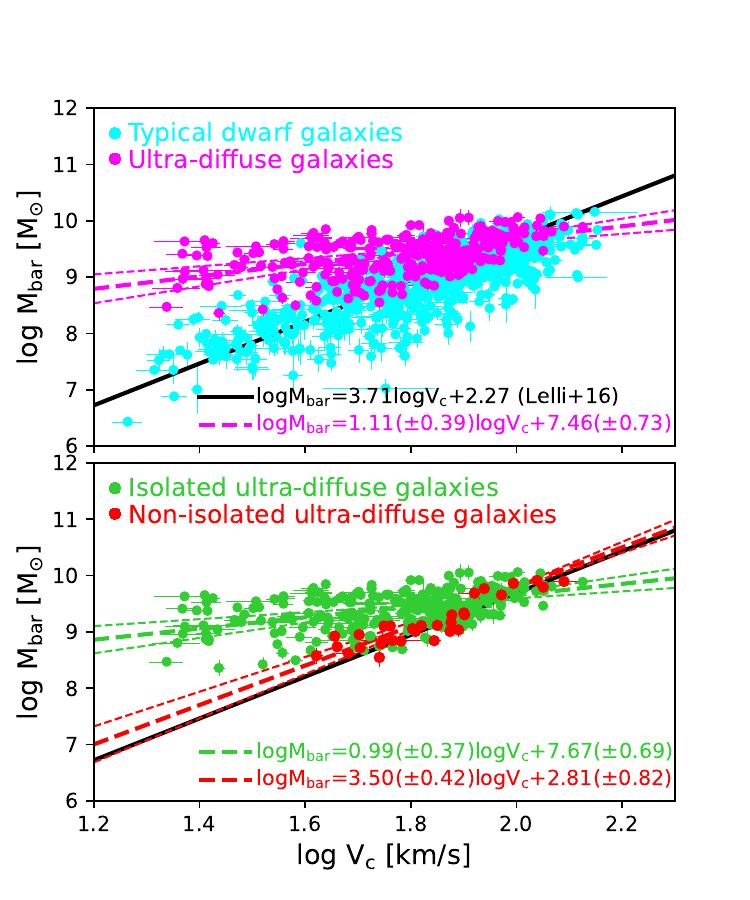} }
 	 \caption{{\bf Upper panel: The bTFRs for HI-bearing UDGs (depicted by magenta circles) and typical dwarf counterparts (depicted by cyan circles)}. The bTFR observed in typical dwarfs \citep{Lelli16} is represented by the black line, while the magenta dashed lines correspond to the best linear fitting results (including the average and 1$\sigma$ uncertainty) for our comprehensive sample of 275 UDGs in this study. {\bf Lower panel: A comparative analysis of the bTFRs between isolated UDGs (depicted by green circles) and non-isolated UDGs (depicted by red circles)}. The dashed lines also represent the optimal linear fitting outcomes for the isolated and non-isolated UDG samples, respectively.
         }
	  \label{relation}
\end{figure}

In the upper panel of Fig.~\ref{relation}, we present the baryonic Tully-Fisher relations (bTFRs) for both UDGs (depicted in magenta) and their dwarf counterparts (depicted in cyan). Consistent with previous investigations into UDG bTFRs \citep{Pina20,Karunakaran20,Hu23}, our UDG sample exhibits a notable departure from the bTFRs observed in typical galaxies \citep{McGaugh00,Lelli16}. Specifically, the UDG bTFR is characterized by a shallower slope of $\log M_{\rm{bar}}=1.11(\pm0.39)\log V_{\rm{c}}+7.46(\pm0.73)$. 

\begin{figure}
	\centerline{ \includegraphics[width=\columnwidth]{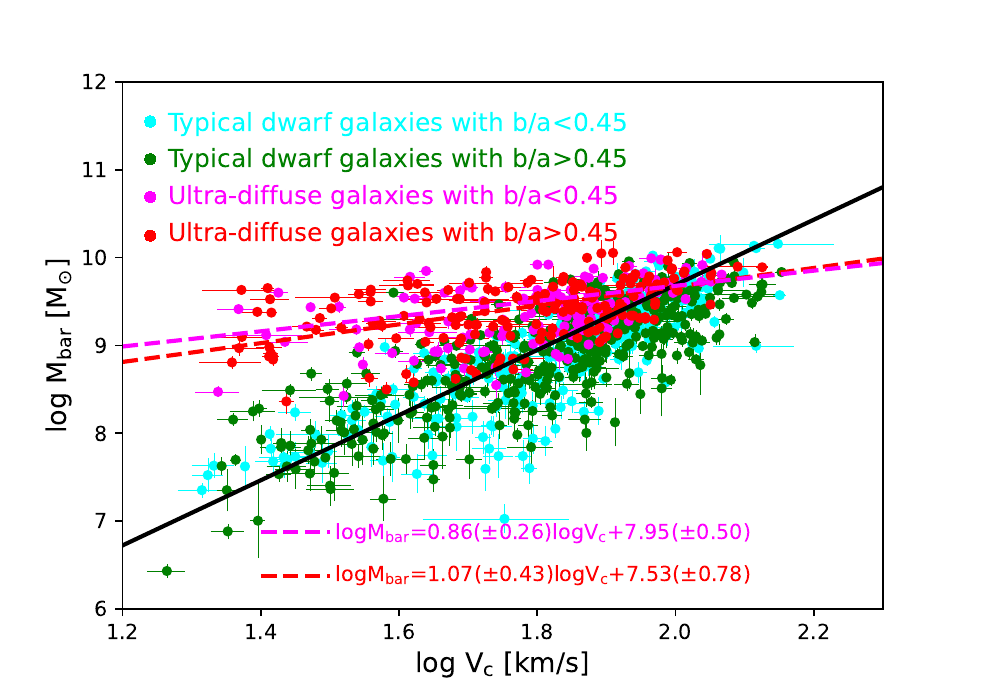} }
	\vspace{-4mm}
 	 \caption{{\bf The bTFRs for UDGs and typical dwarf counterparts with different apparent axis ratios (i.e., different inclinations)}. Similar to the representation in Fig.~\ref{relation}, the two dashed lines correspond to the optimal linear regression results obtained for the UDG subsamples with $b/a<0.45$ and $b/a>0.45$, respectively.
         }
	  \vspace{-4mm}
	  \label{TFR_bar}
\end{figure}

\begin{figure}
	\centerline{ \includegraphics[width=\columnwidth]{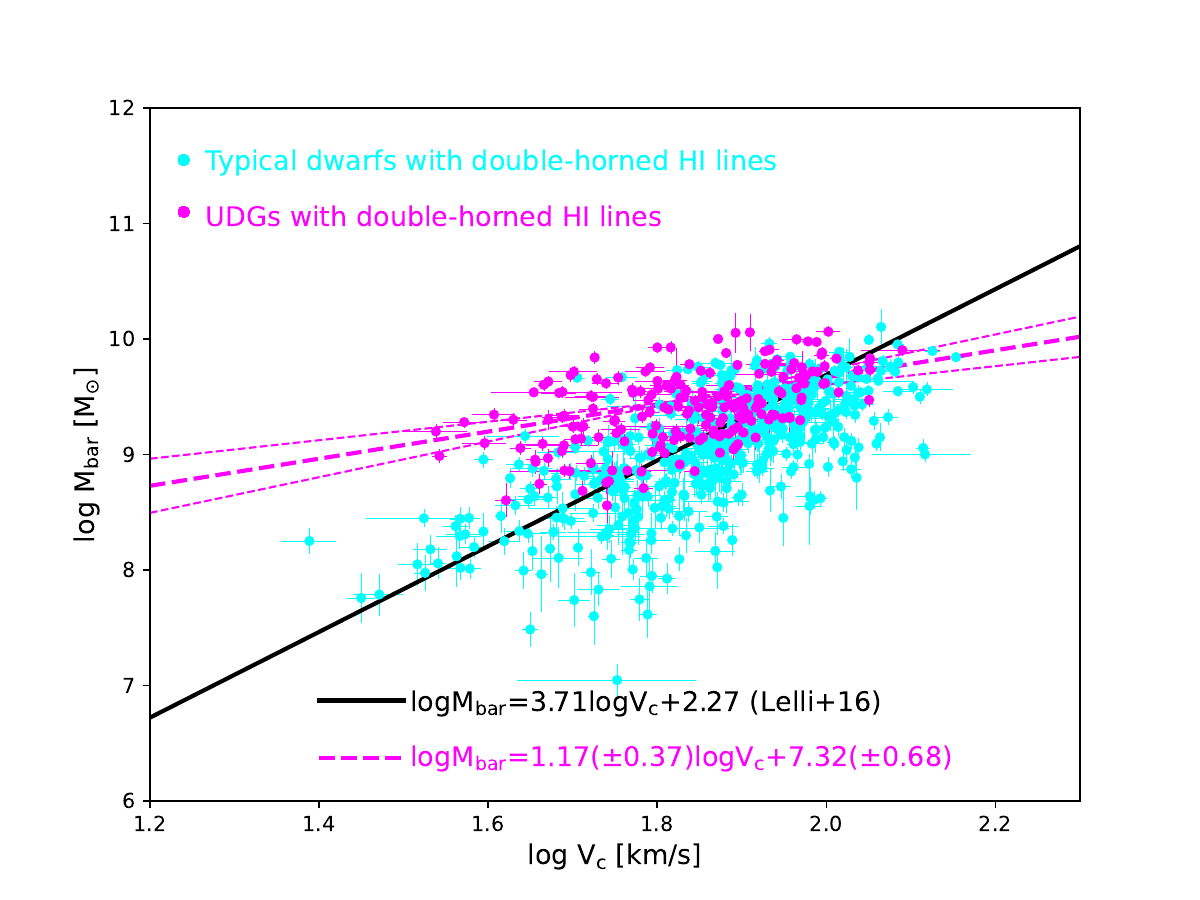} }
	\vspace{-4mm}
 	 \caption{{\bf The bTFRs for UDGs (depicted in magenta) and typical dwarf counterparts (depicted in cyan) exhibiting double-horned HI line profiles}. Similar to the representation in Fig.~\ref{relation}, the dashed lines correspond to the optimal linear regression results obtained for the UDG sample.
         }
	  \vspace{-4mm}
	  \label{dispersion_relation}
\end{figure}

This deviation cannot be ascribed to the possible misalignment in the HI and optical inclination angles, as such misalignment would manifest as a large scatter in $V_{\rm{c}}$ rather than a systematic deviation. Furthermore, it cannot be ascribed to differing inclination angles between UDGs and typical dwarf samples, as the fainter objects detected are more likely to possess larger inclinations. As depicted in Fig.~\ref{TFR_bar}, the bTFRs of UDGs and typical dwarfs exhibit consistent behavior across various inclination angle ranges.

Could the discrepancy be linked to limitations in telescope detection? From the perspective of HI detection, ALFALFA operates as a blind extragalactic HI survey that functions independently of galactic environments and internal properties. Given that the Arecibo beam size is approximately $3.5'$ \citep{Haynes18}, both UDGs and typical dwarfs can be considered as ``point sources'' with $r_{\rm{HI}}\ll 3.5'$. Consequently, UDGs and typical dwarfs should have equivalent HI mass detection limits, resulting in comparable completeness levels for both samples. 

However, the detection completeness of UDGs and typical dwarfs in the SDSS dataset may differ due to the fainter surface brightness of UDGs, particularly those with low stellar masses and extremely large effective radii. Nevertheless, it is worth noting that observations and simulations consistently indicate that the majority (approximately 80\% to 90\%) of UDGs actually possess relatively small effective radii, with $R_{\rm{\star,e}}$ ranging from 1.5 to 3.0~kpc \citep{Yagi16,vanderBurg16,Rong17a,Liao19,DiCintio17}. Therefore, we can estimate the threshold of absolute magnitude beyond which the selection effect does not significantly impact the completeness of UDGs. 

For the SDSS $i$-band, the 3$\sigma$ (1$\sigma$) detection limit of surface brightness is approximately $25.9$ ($27.1$)~mag/arcsec$^2$ \citep{Kniazev04}. A UDG is considered well-detected if the surface brightness at $1.5R_{\star,\rm{e}}$ is $\mu(1.5R_{\star,\rm{e}})\lesssim 27.1$~mag/arcsec$^2$. In this case, we find that the estimates for stellar mass and effective radius of the UDG are minimally affected (e.g., introducing an error of approximately $0.14$ dex in the estimate of stellar mass, which is smaller than the uncertainty in mass and can be neglected). Assuming a S\'ersic profile with an index $n=1$ \citep{Leisman17,Rong20} for the surface brightness distribution of a UDG, we derive the mean surface brightness within the effective radius, $\langle\mu_{\star}\rangle_{\rm{e}}\sim 25.5$~mag/arcsec$^2$, from $\mu(1.5R_{\star,\rm{e}})=27.1$~mag/arcsec$^2$ \citep{Graham05}. Since the $i$-band absolute magnitude of a UDG is given by $M_{i}\simeq \langle\mu_{\star}\rangle_{\rm{e}}-2.5\log(2\pi R_{\star,\rm{e}}^2)-36.57$ \citep{Graham05}, the magnitude threshold for a face-on UDG with $R_{\star,\rm{e}}=3.0$~kpc is $M_i\sim -15.5$~mag. Therefore, for our UDG sample, comprising 98.5\% of UDGs with $M_i<-15.5$ mag, the selection effect is not significant. Beyond this magnitude threshold, only 10\% to 20\% of UDGs have not been detected by SDSS. This implies that even if all of the undetected 10\% to 20\% of UDGs exhibit small $M_{\rm{bar}}$ comparable to typical dwarfs, our conclusions would remain unchanged.

Therefore, the deviation observed in the bTFR of UDGs is not attributed to the detection limits of telescopes.

Could the bTFR discrepancy of UDGs be caused by the possibility of UDGs being dominated by velocity dispersion rather than rotation? Previous work has demonstrated that HI in isolated UDGs tends to be distributed in thin discs, exhibiting regular rotation \citep{Pina20,Li21}. The average ratio of gas velocity dispersion to rotation velocity, denoted as $\langle \sigma /V_{\rm{c}}\rangle$, is found to be less than $20\%$ \citep{Pina20}. 

However, a subset of UDGs may be characterized by a prevalence of velocity dispersion rather than regular rotation. These UDGs are identified by their HI line profiles, which often manifest as a single-horned shape \citep{ElBadry18}. Within our UDG sample, approximately 30\% display these single-horned HI line profiles. To ensure the robustness of our analysis, we exclude these potentially dispersion-dominated UDGs from further investigation, and focuses exclusively on UDGs with double-horned HI line profiles, which are indicative of regular rotation. As detailed in Fig.~\ref{dispersion_relation}, when compared to typical dwarf galaxies, UDGs with double-horned HI line profiles consistently demonstrate a deviated bTFR. Therefore, the deviation observed in the bTFR of UDGs is also not attributable to the possibility that UDGs being dominated by velocity dispersion. 

In conclusion, the deviation of UDG bTFR may appears to have a physical basis.


\subsection{bTFR dependence on environment: UDGs are abundant in baryonic matter, not lacking in dark matter}\label{sec:3.2}

It is worth noting that the divergent bTFR is primarily driven by the low-mass UDGs characterized by small circular velocities ($\log V_{\rm{c}}\sim 1.4\--1.9$), which can be theoretically explained by two plausible formation scenarios. Firstly, these UDGs may have previously been embedded in more massive halos but experienced dark matter loss due to tidal interactions \citep{vanDokkum18,Jing19} or galaxy collisions \citep{vanDokkum}, rendering them dark matter-deficient galaxies. Alternatively, these UDGs may have originated in genuine low-mass halos but have accumulated a greater amount of baryonic matter compared to their typical dwarf counterparts. 

The first scenario suggests that these low-mass UDGs are more likely to reside in high-density environments characterized by stronger tidal forces and an increased likelihood of galaxy encounters. Therefore, we investigate the environments of our UDG sample. In order to examine the environmental characteristics of each galaxy in our sample, we employ the galaxy group and cluster catalog developed by \cite{Saulder16}. This catalog is constructed based on the SDSS DR12 \citep{Alam15} and 2MASS Redshift Survey \citep{Huchra12}, utilizing the friends-of-friends group finder algorithm. Importantly, the study conducted by \cite{Saulder16} comprehensively accounts for various observational biases, such as the Malmquist bias and the `Fingers of God' effect. To determine the environmental classification of galaxies, we adopt the criteria established by \cite{Guo20}. Specifically, galaxies are classified as non-isolated if they are located within a distance of three times the virial radius of the nearest galaxy group or cluster. Conversely, galaxies that do not satisfy this criterion are classified as isolated.

As shown in the lower panel of Fig.~\ref{relation}, we observe that isolated UDGs (depicted in green) exhibit a more pronounced deviation in the bTFR, while UDGs in high-density environments (depicted in red) show a bTFR closer to that of typical dwarfs. This finding supports the notion that the divergent bTFR of UDGs is more likely attributed to the excessive accumulation of baryonic matter in UDGs compared to typical dwarf counterparts, rather than a deficiency in dark matter. Therefore, the formation of UDGs cannot be ascribed to possibility (\Rmnum{2}), i.e., the elevated $V_{\rm{c}}$ and diminished $M_{\star}$.


\subsection{Constrains on UDG formation models}\label{sec:3.3}

\begin{figure*}
	\centerline{ \includegraphics[width=0.9\textwidth]{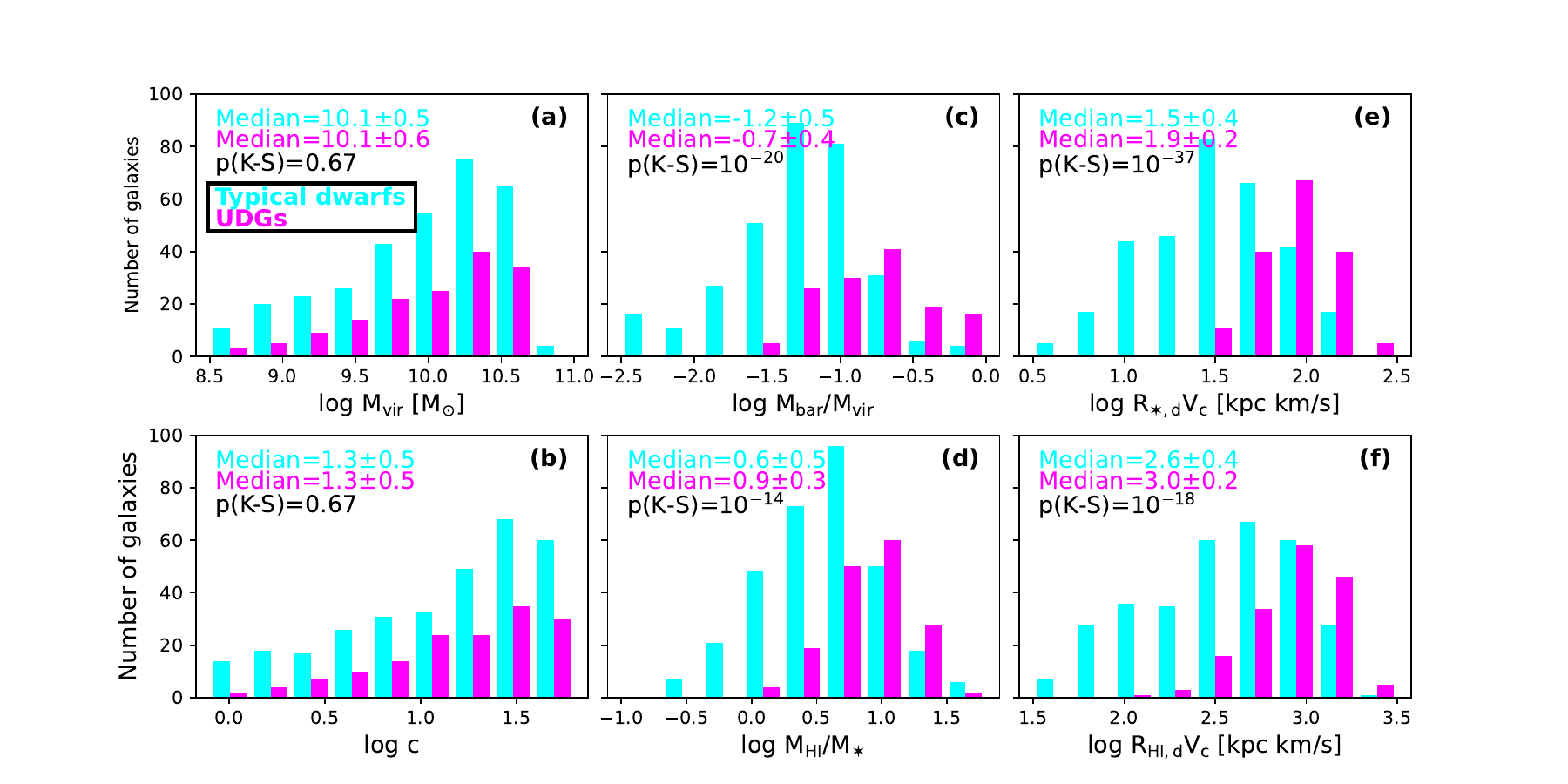} }
	\vspace{-4mm}
 	 \caption{{\bf The comparative analysis between isolated UDGs (depicted in magenta) and typical dwarf counterparts (depicted in cyan).} Panels~a to f present histograms displaying the distributions of halo masses, halo concentrations, baryonic mass fractions, HI-to-stellar mass ratios, as well as $R_{\star,\rm{d}}V_{\rm{c}}$ and $R_{\rm{HI,d}}V_{\rm{c}}$ (which can be considered as proxies for their stellar and HI specific angular momenta), for UDGs and typical dwarfs, respectively. The logarithmic scale is employed to represent the median value of each property, and the Kolmogorov-Smirnov test $p$-value is provided to assess the disparity in property distributions between the two galaxy samples, as indicated in the corresponding panel.
         }
	  \vspace{-4mm}
	  \label{compare}
\end{figure*}

\begin{figure*}
	\centerline{ \includegraphics[width=0.9\textwidth]{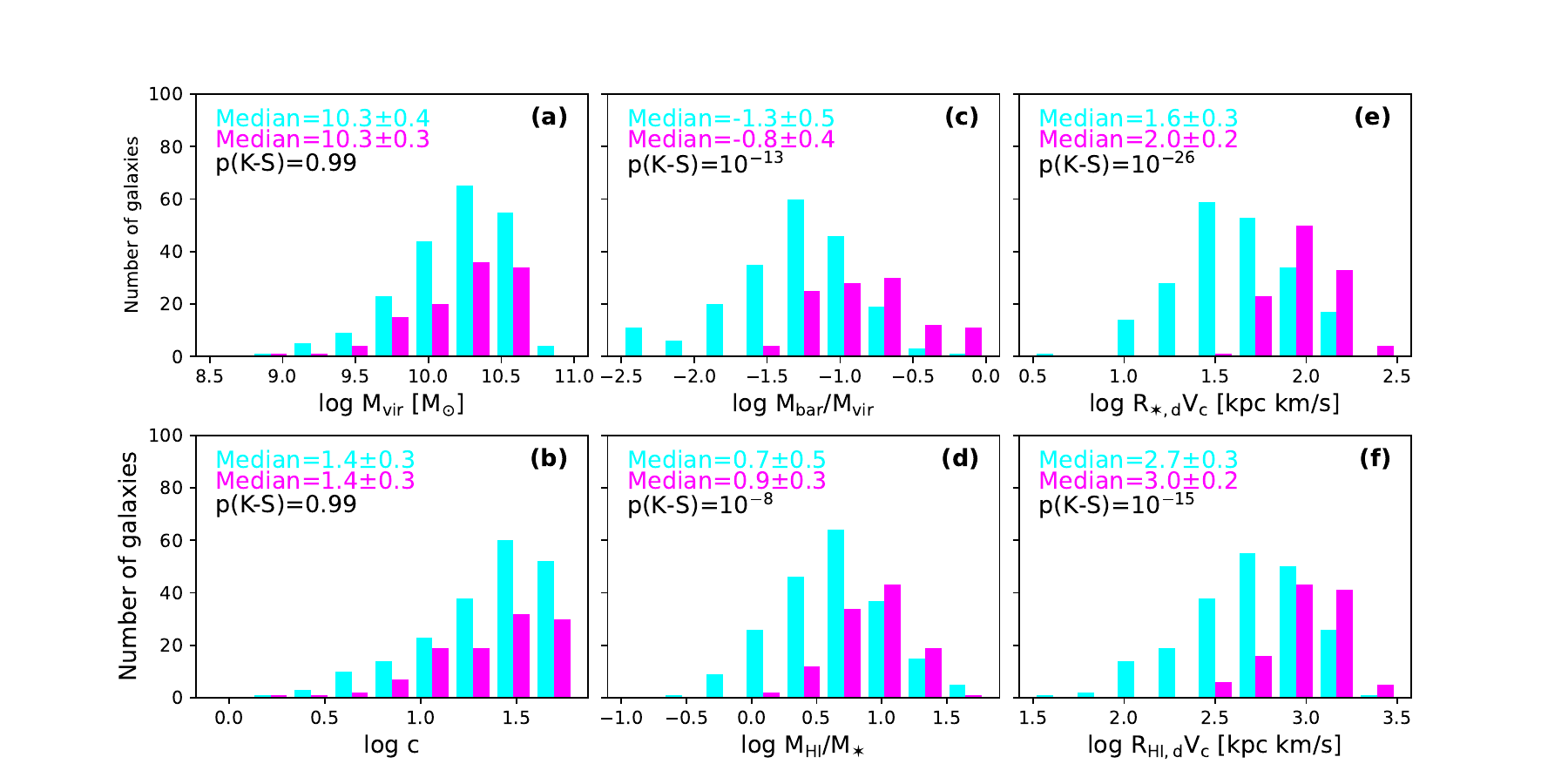} }
	\vspace{-4mm}
 	 \caption{{\bf A comparative analysis between isolated UDGs (represented in magenta) and typical dwarf counterparts (represented in cyan) exhibiting double-horned HI line profiles}. The selected samples encompass the same range of circular velocities with $1.5\lesssim \log V_{\rm{c}} \lesssim 1.9$. Each panel within this study is analogous to the corresponding one depicted in Fig.~\ref{compare}.
         }
	  \label{dispersion_compare}
\end{figure*}

We also directly estimate the halo masses $M_{\rm{vir}}$ for both isolated UDGs and typical dwarf counterparts with the method described in section~\ref{sec:2.2}. Our findings reveal that the maximum halo mass in our UDG sample is approximately $2\times 10^{11}\ \rm M_{\odot}$, considerably less massive than the dark matter halo of an L$^{\star}$ galaxy, which typically reaches $10^{12}\ \rm M_{\odot}$. This discrepancy also contradicts the failed L$^\star$ formation model proposed for UDGs \citep{vanDokkum15} (possibility \Rmnum{2}). 

Subsequently, we conduct a direct comparison of the halo masses and baryonic fractions ($M_{\rm bar}/M_{\rm vir}$) between the isolated UDG and typical dwarf samples, specifically within the circular velocity range of $1.4\lesssim \log V_{\rm{c}} \lesssim 1.9$. As illustrated in Fig.~\ref{compare}, while the distributions of halo masses exhibit similarity between the two samples, the baryonic mass fractions and HI-to-stellar mass ratios ($M_{\rm HI}/M_{\star}$) of UDGs are significantly higher, indicating an excess of baryonic matter in UDGs, predominantly in the form of gas.

In empirical galaxy formation models, similar baryonic mass fractions are expected in isolated dark matter halos with comparable masses. Therefore, the higher baryonic fractions observed in UDGs suggest an ineffective outflow driven by supernovae, which fails to expel gas from halos as effectively as in typical dwarf galaxies. In other words, the heightened abundance of HI gas in UDGs cannot be adequately attributed to significant alterations of gravitational potentials \citep{Mo04} (possibility \Rmnum{3}). The gravitational potentials of the isolated UDGs and typical dwarf counterparts, characterized by the concentrations ($c$) of their dark matter halos, are statistically comparable, as indicated in panel~(b) of Fig.~\ref{compare}. Consequently, the more compelling explanation for the emergence of extended stellar disk radii and reduced stellar densities in UDGs lies in their elevated specific angular momenta $S_{\star}$, i.e., (possibility \Rmnum{1}).

\subsection{Higher specific angular momentum in UDGs}\label{sec:3.4}

Hence, we proceed to estimate and compare the specific angular momenta of the stellar and HI-gas components for the two samples. For a stellar disk, its specific angular momentum can be determined using equation~(\ref{eq1})
\begin{equation} 
	S_{\star}\propto R_{\star,\rm{d}} V_{\rm{c}}.
	 \label{sam_star} 
\end{equation}
Here the scale length is approximately $R_{\star,\rm{d}}\simeq R_{\star,\rm{e}}/1.678$ for an exponential stellar disk profile \citep{Graham05}. Similarly, the specific angular momentum of an HI-gas disk can be expressed as
\begin{equation} 	
	S_{\rm{HI}}\equiv\frac{J_{\rm{HI}}}{M_{\rm{HI}}}\propto R_{\rm{HI,d}} V_{\rm{c}},
	\label{sam_HI} 
\end{equation}
where $J_{\rm{HI}}$ represents the angular momentum of the HI disk, and the scale length of the HI disk, $R_{\rm{HI,d}}$, can be derived by assuming a relatively thin gas disk in centrifugal balance \citep{Mo98}, characterized by an exponential surface density profile, as
\begin{equation}
	\Sigma_{\rm{HI}}(R)=\Sigma_{{\rm{HI}},0} {\rm{exp}}(-R/R_{{\rm{HI,d}}}),
\end{equation}
where $\Sigma_{{\rm{HI}},0}$ represents the central surface density of the HI disk. The total HI mass $M_{\rm{HI}}$ is related to the scale length as
\begin{equation} M_{\rm{HI}} = 2 \pi \Sigma_{{\rm{HI}},0} R_{{\rm{HI,d}}}^2 \label{HIeq_mass}. \end{equation} 
Furthermore, by definition, at $r_{\rm{HI}}$, we have 
\begin{equation} \Sigma_{{\rm{HI}},0} {\rm{exp}}(-r_{\rm{HI}}/R_{{\rm{HI,d}}})=1\ \rm  M_{\odot}\rm{pc^{-2}}. \label{HIeq_3} \end{equation} 
Using equations~(\ref{HIeq_mass}) and (\ref{HIeq_3}), we can calculate the value of $R_{{\rm{HI,d}}}$ for each galaxy in our sample.
	
Therefore, by utilizing equations~(\ref{sam_star}) and (\ref{sam_HI}), the comparison of the stellar and HI specific angular momenta between the UDG and dwarf counterpart samples can be simplified as a comparison of the observable quantities $R_{\star,\rm{d}}V_{\rm{c}}$ and $R_{\rm{HI,d}}V_{\rm{c}}$, respectively. As illustrated in panels~(e) and (f) of Fig.~\ref{compare}, the average specific angular momenta (or spins) of UDG stellar and HI-gas disks are approximately 0.4~dex higher than those of the dwarf counterparts on a logarithmic scale. 

We further present the comparative outcomes for the resilient subsets of solitary UDGs and typical dwarfs possessing double-horned HI profiles. As illustrated in Fig.~\ref{dispersion_compare}, the findings align with those of the entire dataset.

\section{Discussion and Summary}\label{sec:4}

In this investigation, we initially classified the existing UDG formation models into three possibilities, (\Rmnum{1}) higher $S_{\star}$, (\Rmnum{2}) elevated $V_{\rm{c}}$ and diminished $M_{\star}$, and (\Rmnum{3}) a change in gravitational potential. Subsequently, we selected the sample of HI-bearing UDGs in observational data and compared their properties with those of typical dwarf counterparts. Our analysis ultimately dismisses the second and third formation possibilities, attributing the origin of HI-bearing UDGs to their elevated specific angular momentum, predominantly aligning with possibility (\Rmnum{1}). Nonetheless, we are unable to definitively discern whether the higher $S_{\star}$ of UDGs arises from a higher halo spin or a greater spin of the accreted gas, albeit with comparable halo spin to typical dwarfs.

Although this formation model is not novel and has been supported by cosmological hydrodynamical simulations \citep{Benavides23}, as well as semi-analytic galaxy formation models implemented on $N$-body simulations \citep{Rong17a,Amorisco16}, and also been proposed by observational UDG studies \citep[e.g.,][]{Pina20}, our study, for the first time, provides statistical observational evidence to validate this formation mechanism and reject other formation possibilities.

As mentioned in section~\ref{sec:1}, several observational studies on the kinematics of UDGs do not support the higher $S_{\star}$ compared to typical dwarfs. However, it is important to note that these observations primarily pertain to UDGs in high-density environments, which may have undergone tidal heating, possibly leading to a reduction in their $S_{\star}$.

The discovery of higher $S_{\star}$ and $S_{\rm{HI}}$ in UDGs provides a comprehensive understanding of UDG formation. We propose that UDGs emerged within dark matter halos possessing masses akin to those of dwarf galaxies. Nevertheless, in contrast to typical dwarfs, the halos of UDGs exhibited augmented rotational velocities or experienced the acquisition of faster rotating circumgalactic medium from their surroundings. As a consequence, UDGs harbor highly-spinning HI gas. This highly-spinning gas demonstrates greater resilience against angular momentum loss, impeding its descent towards the galactic core and subsequent condensation and cooling necessary for providing the cold ($<100\ \rm{K}$) fuel essential for star formation in the central regions of the halo \citep{Peng20}. The inflow of gas into the central region of a UDG is therefore gradual and continuous, ensuring a sustained supply. This gradual inflow precludes the instantaneous ignition of a substantial burst of star formation or the triggering of supernova explosions en masse, thus preventing the expulsion of a large amount of HI gas from UDGs beyond the confines of the dark matter halo. This intricate mechanism ultimately culminates in the accumulation of excessive HI masses, reduced star formation efficiencies, diminished stellar densities, and the expansion of stellar disk radii observed in UDGs. 

Many observational and theoretical studies on UDGs concur that these low-surface-brightness objects tend to have lower star formation rates and a more extended star formation history \citep[e.g.,][]{Rong20,Trujillo17,DiCintio17}, which conversely support the higher $S_{\rm{HI}}$ in UDGs based on the aforementioned scenario. Conversely, the protracted star formation history may also account for the abundant presence of globular clusters in numerous UDGs \citep{Forbes20}. If star formation predominantly occurred in the early Universe, the constituent globular clusters may have undergone substantial mass loss due to stellar evolution \citep{Lamers10,Carretta10,Decressin18}, rendering them faint and challenging to detect. In contrast, an extended star formation process naturally results in a broad range of ages among member globular clusters, potentially preserving a larger number of detectable globular clusters at redshift $z\sim 0$.

\begin{acknowledgments}

Y.R. acknowledges supports from the NSFC grant 12273037, and the CAS Pioneer Hundred Talents Program (Category B), as well as the USTC Research Funds of the Double First-Class Initiative. W.D. is supported by the Youth Innovation Promotion Association, Chinese Academy of Sciences No.~2020057. Q.G. is supported by the National SKA Program of China No. 2022SKA0110201, the CAS Project for Young Scientists in Basic Research Grant No. YSBR-062, and the European Union's Horizon 2020 Research and Innovation Programme under the Marie Sk\l odowska-Curie grant agreement No.~101086388. H.Y.W. is supported by CAS Project for Young Scientists in Basic Research, Grant No. YSBR-062, and the NSFC grant 12192224. H.X.Z. acknowledges support from the NSFC grant 11421303.

\end{acknowledgments}




\end{document}